\documentclass[fleqn,usenatbib,useAMS]{mnras}

\usepackage{graphicx}	
\usepackage{amsmath}	
\usepackage{amssymb}	
\usepackage{multicol}        
\usepackage{bm}		
\usepackage{pdflscape}	
\usepackage{lscape}
\usepackage[flushleft]{threeparttable}

\usepackage[T1]{fontenc}
\usepackage{ae,aecompl}

\usepackage{newtxtext,newtxmath}

\title[Characterization of W75N~(B)]{\textit{Characterizing the radio continuum nature of sources in the massive star-forming region W75N~(B)}}

\author[A. Rodr\'{\i}guez-Kamenetzky]{A. Rodr\'{\i}guez-Kamenetzky,$^{1}$\thanks{Contact e-mail: \href{mailto:adriana.rodriguez@unc.edu.ar}{adriana.rodriguez@unc.edu.ar}} 
C. Carrasco-Gonz\'alez,$^2$ J. M. Torrelles,$^{3,4}$
\newauthor W. H. T. Vlemmings,$^5$
L. F. Rodr\'iguez,$^2$ 
G. Surcis,$^6$ J. F. G\'omez,$^7$ 
J. Cant\'o,$^8$ 
\newauthor C. Goddi,$^{9,10}$ 
J.  S. Kim,$^{11}$ S. -W. Kim,$^{11}$
N. A\~nez-L\'opez,$^{3,4}$ S. Curiel$^8$ 
\newauthor and H. J. van Langevelde$^{12, 13}$
\\
\scriptsize{$^{1}${Instituto de Astronom\'ia Te\'orica y Experimental, (IATE-UNC), X5000BGR C\'ordoba, Argentina}}\\
\scriptsize{$^{2}${Instituto de Radioastronom\'ia y Astrof\'isica (IRyA-UNAM), 58089 Morelia, M\'exico}}\\
\scriptsize{$^{3}${Institut de Ci\`encies de l'Espai (ICE, CSIC), Can Magrans s/n, E-08193, Cerdanyola del Vall\`es, Spain}}\\
\scriptsize{$^{4}${Institut d'Estudis Espacials de Catalunya (IEEC), E-08034, Barcelona, Spain}}\\
\scriptsize{$^{5}${Department of Earth and Space Sciences, Chalmers University of Technology, SE-43992 Onsala, Sweden}}\\
\scriptsize{$^{6}${INAF-Osservatorio Astronomico di Cagliari, Via della Scienza 5, 09047 Selargius (CA), Italy}}\\
\scriptsize{$^{7}${Instituto de Astrof\'{\i}sica de Andaluc\'{\i}a, CSIC, Glorieta de la Astronom\'{\i}a s/n, E-18008 Granada, Spain}}\\
\scriptsize{$^{8}${Instituto de Astronom\'ia, Universidad Nacional Aut\'onoma de M\'exico (UNAM), Apartado Postal 70-264, DF 04510 M\'exico}}\\
\scriptsize{$^{9}${ALLEGRO/Leiden Observatory, Leiden University, P.O. Box 9513, 2300 RA Leiden, The Netherlands}}\\
\scriptsize{$^{10}${Department of Astrophysics/IMAPP, Radboud University Nijmegen, P.O. Box 9010, 6500 GL Nijmegen, The Netherlands}}\\
\scriptsize{$^{11}${Korea Astronomy and Space Science Institute, 776 Daedeokdaero, Yuseong, Daejeon 305-348, Republic of Korea}}\\
\scriptsize{$^{12}${Joint Institute for VLBI ERIC (JIVE), Oude Hoogeveensedijk 4, 7991 PD Dwingeloo, The Netherlands}}\\
\scriptsize{$^{13}${Leiden Observatory, Leiden University, PO Box 9513, 2300 RA, Leiden, The Netherlands}}}

\begin{document}
\label{firstpage}
\pagerange{\pageref{firstpage}--\pageref{lastpage}}
\maketitle

\begin{abstract}
{The massive star-forming region W75N~(B) is thought to host a cluster of massive protostars (VLA~1, VLA~2, and VLA~3) undergoing different evolutionary stages. In this work, we present radio continuum data with the highest sensitivity and angular resolution obtained to date in this region, using the VLA-A and covering a wide range of frequencies  (4-48~GHz), which allowed us to study the morphology and the nature of the emission of the different radio continuum sources. We also performed complementary studies with multi-epoch VLA data and ALMA archive data at 1.3 mm wavelength. We find that VLA~1 is driving a thermal radio jet at scales of $\approx$0.1 arcsec ($\approx$130 au), but also shows signs of an incipient hyper-compact HII region at scales of $\lesssim$ 1 arcsec ($\lesssim$ 1300~au). VLA~3 is also driving a thermal radio jet at scales of  a few tenths of arcsec (few hundred of au). We conclude that this jet is shock-exciting the radio continuum sources Bc and VLA~4 (obscured HH objects), which show proper motions moving outward from VLA~3 at velocities of $\approx$112--118~km/s. We have also detected three new weak radio continuum sources, two of them associated with millimeter continuum cores observed with ALMA, suggesting that these two sources are also embedded YSOs in this massive star-forming region.}
\end{abstract}




\section{Introduction}

Although it is well-known that the most massive stars have a great impact on the galactic environment, many aspects related to their early evolutionary stages still remain unknown. For instance, massive protostars are deeply embedded in dense molecular gas, located at typical distances of few thousand parsecs. Thus, detailed studies of these objects require observations with very high sensitivity and angular resolution. One of the best known massive star-forming regions is W75N~(B), located in the Cygnus X complex at a distance of 1.3~kpc \citep{rygl2012},  comprising dense molecular clouds \citep{dickel1978,persi2006} and showing strong maser emission of different molecular species \citep[e.g.,][]{baart1986,hunter1994,torrelles1997,surcis2009,krasnov2015,colom2018}. 
This region constitutes an excellent laboratory to study early stages of massive star formation, since it hosts a cluster of massive protostars \citep[e.g.,][]{shepherd2003}, probably undergoing different evolutionary phases \citep{torrelles1997}.

Since its discovery, W75N(B) has been widely studied, 
revealing the presence of five radio continuum sources \citep[named VLA~1, VLA~2, VLA~3, VLA~4, and Bc; e.g.,][]{hunter1994,torrelles1997,carrasco2015} and a large-scale high-velocity molecular outflow \citep[e.g.,][]{davis1998,shepherd2003}. Among the five radio continuum sources, VLA~1 was proposed to be an evolved young stellar object (YSO), whereas VLA~2 is probably the least evolved YSO in the region \citep[e.g.,][]{torrelles1997}. These two sources are the only ones in the region that are associated with 22 GHz water \citep[e.g.,][]{torrelles1997,surcis2009, surcis2011,surcis2014,kim2013} and 6.7 GHz methanol maser emission, which was actually detected from a location in between them \citep[e.g.,][]{minier2000, surcis2009}. Furthermore, polarimetric maser observations show the presence of a magnetic field oriented in the direction of the molecular outflow \citep[e.g.,][]{huta2002,surcis2009, surcis2011,surcis2014}. However, despite the deep studies conducted
so far towards W75N(B), the nature of some of the radio continuum sources in the region is not well known yet.

In this work, we analyze radio continuum data obtained with the Karl Jansky Very Large Array (VLA) over a wide range of frequencies (4 to 48~GHz), which provide images with the highest sensitivity (rms = 8~$\mu$Jy/beam) and angular resolution ($0\farcs12\times0\farcs09$, PA= -69$^{\circ}$) obtained to date in this region. Part of these data were presented by \cite{carrasco2015}, who focused their attention on the remarkable source VLA~2, reporting through radio continuum and H$_2$O maser observations the transition from an uncollimated outflow to a collimated outflow over a period of only 18 years.  In this work, we focus on the remaining sources in the field: VLA 1, VLA 3, VLA 4, and Bc (see Fig. \ref{w75n}a). These observations allow us to perform a deep multifrequency study of the morphology of the sources and of their nature. We also analyze Atacama Large Millimeter Array (ALMA) 1.3~mm continuum and spectral line archive data obtained toward this region.

\section{Observations}
\subsection{VLA}

\begin{figure*}
    \centering
    \includegraphics[width=\textwidth]{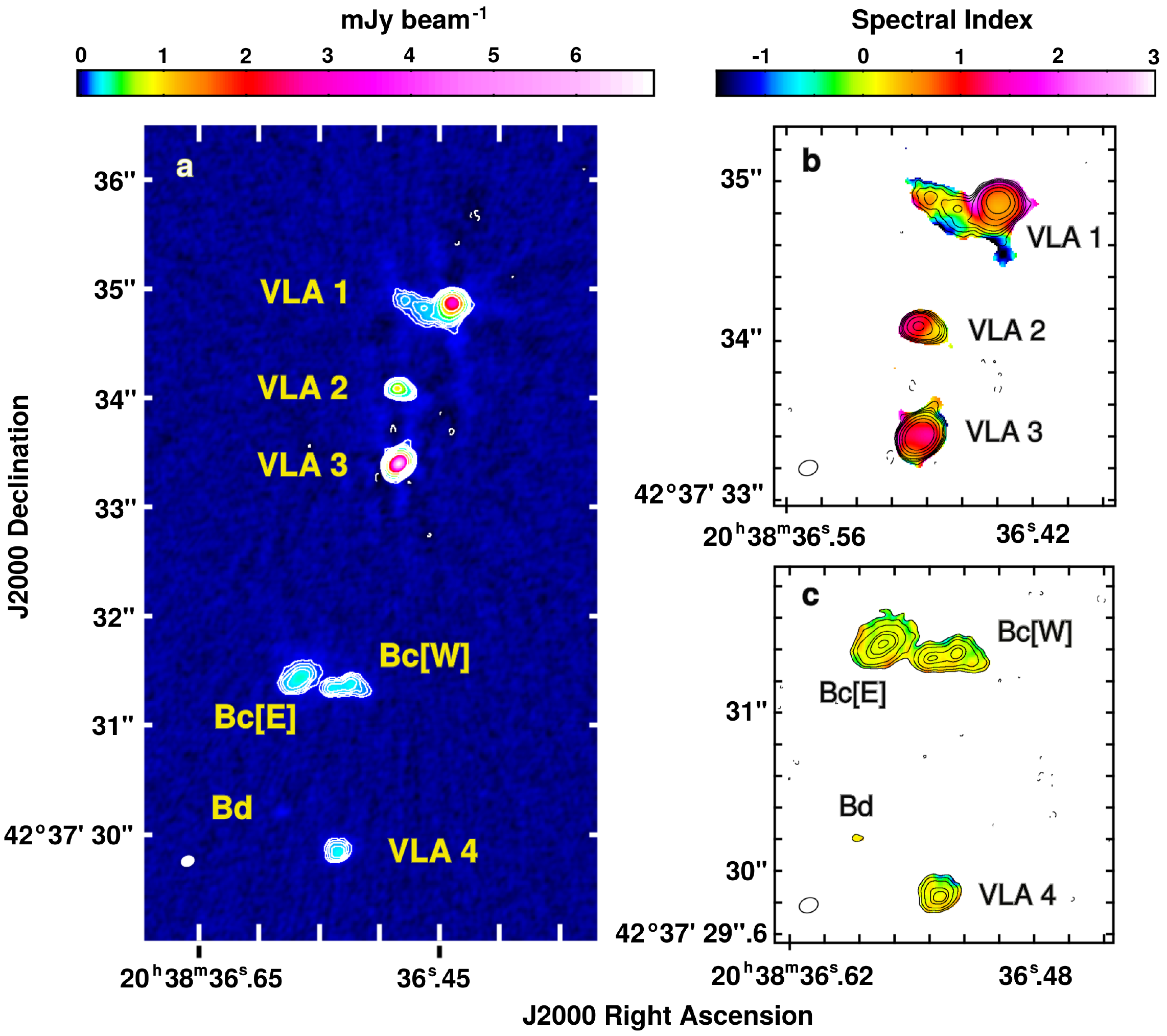}
    \caption{Radio continuum image and spectral index maps of W75N~(B) made by the combination of C, Ku, K, and Q bands (epoch 2014.29), using multifrequency synthesis cleaning and Briggs weighting (robust 0). (a) Continuum image: contours are --4, 9, 13, 18, 25, 50, 100, and 200 times the rms, 8~$\mu$Jy/beam. Panels (b) and (c) show a close-up of the northern region containing sources VLA 1, 2, and 3, and the southern region containing sources Bc and VLA 4, respectively. In both cases, intensity contours of panel (a) are shown over the spectral index map (color scale). The pixels shown in spectral index maps are those with S/N > 7 in the continuum image. Synthesized beam = 0$\farcs$12$\times$0$\farcs$09 (PA = --69$^{\circ}$).
    }\label{w75n}
\end{figure*}

\begin{figure*}
    \centering
    \includegraphics[width=\textwidth]{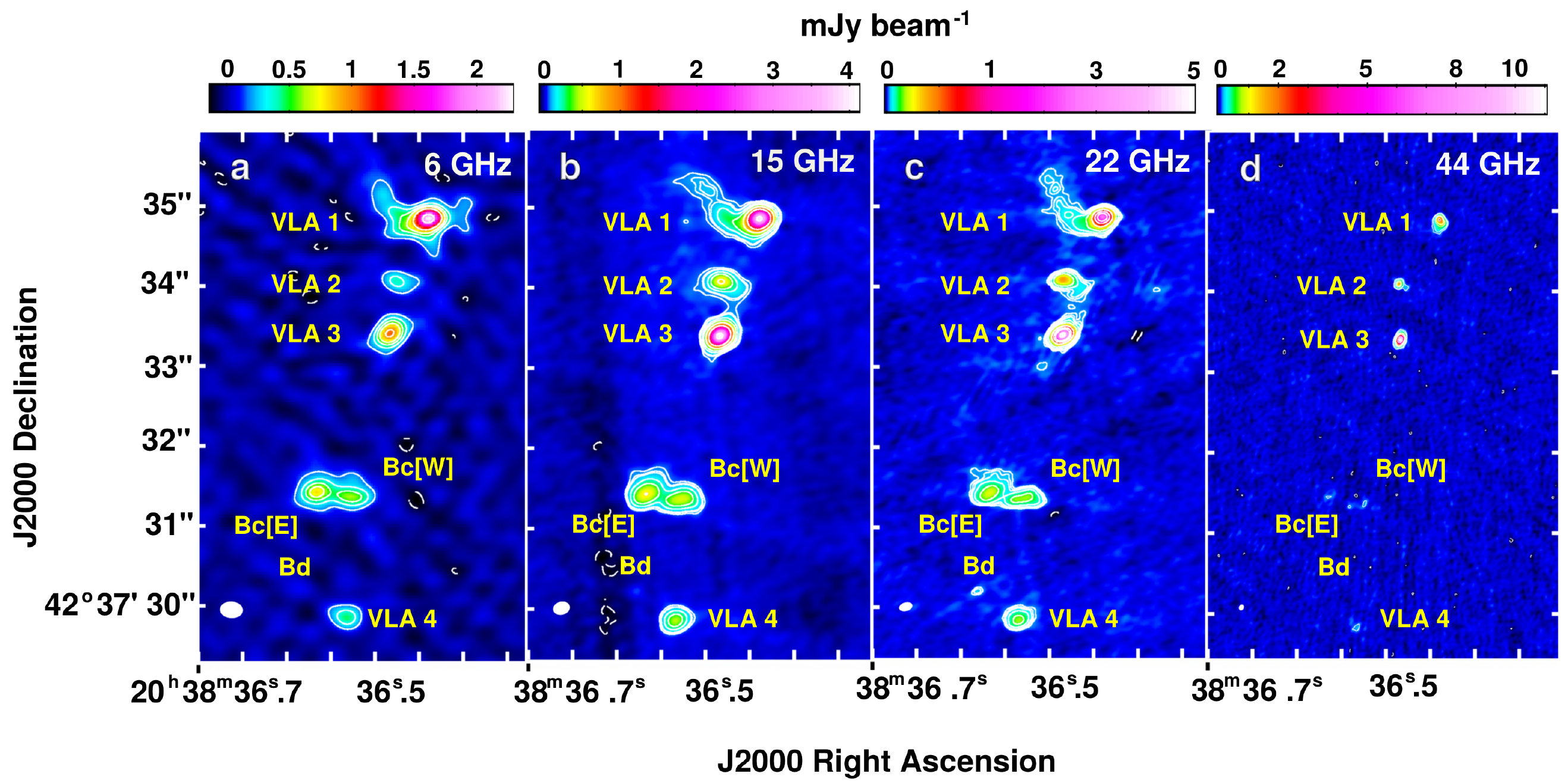}
    \caption{Radio continuum images of W75N~(B) at C (6~GHz), Ku (15~GHz), K (22~GHz), and Q (44~GHz) bands (epoch 2014.29) are shown in panels a, b, c, and d, respectively. White dashed and solid contours represent negative and positive values, respectively, corresponding to different sigma levels: -3, 5, 10, 15, 20, 30, 60 times 30~$\mu$Jy/beam (6~GHz image, uniform weighting); --4, 5, 7, 8, 15, 25, 50, 100, 300 times 10~$\mu$Jy/beam (15~GHz image, natural weighting); --4, 5, 7, 9, 15, 30, 100, 200, 300, 600 times 10~$\mu$Jy/beam (22~GHz image, natural weighting); --3, 5, 10, 30, 50, 100, 300, 500 times 20~$\mu$Jy/beam (44~GHz image, natural weighting). In each panel, the synthesized beam is indicated with a white ellipse at the bottom left.
    }\label{bands}
\end{figure*}

The star-forming region W75N~(B) was observed with the VLA of the National Radio Astronomy Observatory (NRAO)\footnote{NRAO is a facility of the National Science Foundation operated under cooperative agreement by Associated Universities, Inc.} in its A-configuration at C (6 cm), Ku (2 cm), K (1.3 cm), and Q (7 mm) bands (project code 14A-007).
A detailed description of the observations and calibration procedures can be found in \citet{carrasco2015}. Deconvolved images were obtained with the task CLEAN of the Common Astronomy Software Applications (CASA\footnote{https://science.nrao.edu/facilities/vla/data-processing}, version 4.1.0) data reduction package, using multifrequency synthesis (parameter nterms = 2) and multi-scale cleaning \citep{rau2011}.  Primary beam corrections were applied. We split each band data set to build images of narrower bandwidth (1 and 2~GHz), using different weightings, i.e., natural, uniform, and Briggs \citep{briggs} to achieve the best compromise between sensitivity and angular resolution, depending on the analysis performed (e.g., spectral energy distributions, angular size vs. frequency). We also made a single image combining all four bands (C, Ku, K, and Q; Fig. \ref{w75n}), as well as individual images integrating the full bandwidth of each frequency band (Fig. \ref{bands}). Moreover, the multifrequency synthesis cleaning technique allows us to obtain a spectral index map covering the entire range of the observed frequencies.

\begin{figure*}
    \centering
    \includegraphics[width=\textwidth]{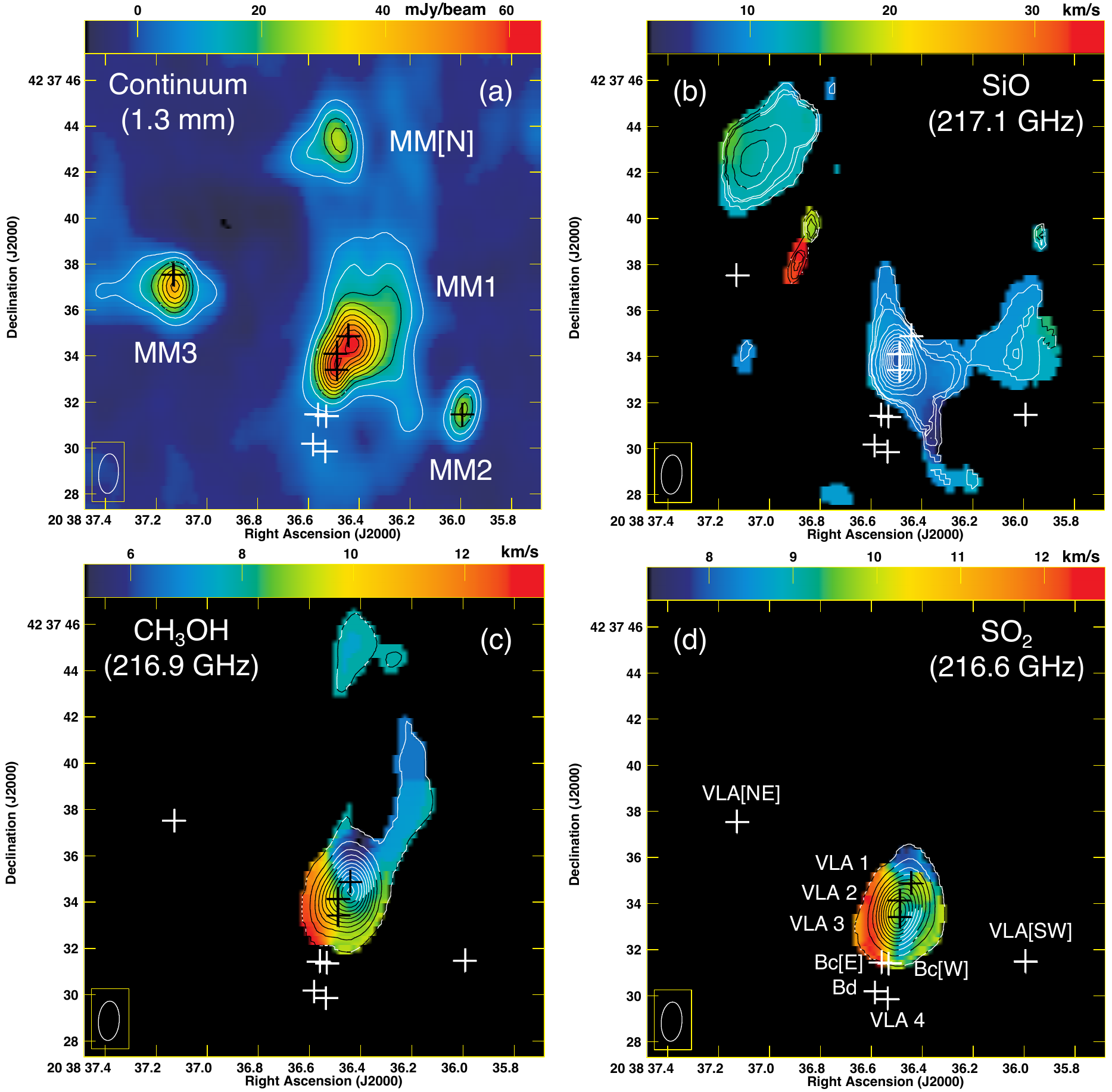}
    \caption{(a) Colour image and contour map of the continuum emission at 1.3 mm.  Contour levels are 1, 2, 3, 4, 5, 6, 7, 8, 9, 10 $\times$ 6.1~mJy~beam$^{-1}$. The mm sources MM1, MM2, and MM3 are labeled \citep[nomenclature by][]{minh2010}. A new mm source detected with ALMA $\sim$9~arcsec north from MM1 is labeled as MM[N]. Crosses indicate the positions of the radio continuum sources at cm wavelengths identified with the VLA (this work). (b) Contour map of the integrated intensity and velocity field colour image (moment of order 1) of the SiO (217.1~GHz) line. Contour levels are 0.4, 0.8, 1, 2, 3, 4, 5, 6, 7, 8, 9, 10$\times$0.27~Jy~beam$^{-1}$~km~s$^{-1}$. (c) Same as the previous panel but for the CH$_3$OH (216.9 GHz) line. Contour levels are 0.1, 1, 2, 3, 4, 5,6, 7, 8, 9, 10$\times$0.6~Jy~beam$^{-1}$~km~s$^{-1}$. (d) Same as the previous panel but for the SO$_2$ (216.6 GHz) line. Contour levels are 0.1, 1, 2, 3, 4, 5, 6, 7, 8, 9, 10$\times$0.8~Jy~beam$^{-1}$~km~s$^{-1}$. Primary beam corrections have not been applied to these images, given that the mm cores are detected at the edge of the ALMA primary beam (FWHM $\simeq$ 27~arcsec; see Section \ref{alma-sources}). Assuming that the ALMA primary beam can be approximated by a Gaussian function, the intensities given in these images should be corrected by factors of $\sim$4, 10, 2, and 1.4 at the positions of MM1, MM2, MM3, and MM[N], respectively.
    }\label{figualma}
\end{figure*}

All the radio continuum images presented in this paper, as well as the spectral energy distribution analysis of the different sources are based on the VLA project code 14A-007 (epoch 2014.29). However, in order to study the kinematics of some of the sources in the region  (VLA~4 and Bc) we also reanalyzed previously reported, multi-epoch VLA archive data \citep[project codes AT141, AF381, and AS831; see][for details on the observations]{carrasco2010}. This, along with our new K-band observations, allow us to compute proper motions in a period spanning 22 years, from 1992 to 2014. Calibration of these archive data was undertaken following standard VLA procedures, using the Astronomical Image Processing System (AIPS\footnote{http://www.aips.nrao.edu/index.shtml}) data reduction package.

Parameters of the data sets and images are summarized in Table \ref{tbl-data} and Table \ref{tbl-images}, respectively.

\subsection{ALMA}

W75N~(B) was observed with ALMA at 1.3~mm during three sessions, on May 6th, 7th, and 11th 2018 (archive ALMA data, project code: 2017.1.01593.S). In total, approximately 16 minutes were spent on source. During the session on May 7th, only one minute of useful data on W75N (B) was obtained. The phase center for the W75N~(B) observations was RA(J2000) = 20:38:37.0 and Dec(J2000) = +42:37:51.0, which is $\sim$18~arcsec north of the VLA~1-VLA~2-VLA~3 sources. As a result, the mm continuum sources discussed in this work are detected towards the edge of the ALMA primary beam (FWHM $\simeq$ 27~arcsec). The ALMA images presented in this paper (Section \ref{alma-sources}) have not been corrected by primary beam but they are of good enough quality for the identification of different mm continuum sources. The observations were performed using four spectral windows (spws). Two spws had 1.875~GHz bandwidth and were centered on $217.117$ and $230.552$~GHz. Two further spws had 117.188~MHz bandwidth and were centered at 216.124 and 231.334~GHz. All spws had 1920 spectral channels after hanning smoothing. During the observations, 46 ALMA telescopes participated, with a minimum baseline length of 15~m and a maximum baseline length of 500~m. This resulted in a maximum recoverable scale of $\sim$7 arcsec.

The observations were initially calibrated using the ALMA pipeline of CASA 5.1.1 \citep{mullin2007}.
Subsequently, after excluding the channels showing line emissions, we performed two interactions of phase self-calibration on the continuum emission by using CASA 5.1.1. This improved the continuum signal-to-noise ratio by a factor of three. Next, the data were imaged and cleaned (task TCLEAN) centered on VLA~2, using Briggs weighting with a robust parameter of $0.5$, yielding a synthesized beam size for the 1.3~mm continuum observations of $1$\farcs$73\times0$\farcs$86$ with a position angle of $-4^\circ$ (Fig. \ref{figualma}).

 \begin{table*}
 \center
  \caption{PARAMETERS OF THE VLA OBSERVATIONS}
  \label{tbl-data}
    \begin{threeparttable}
  \begin{tabular}{lccccc}
    \hline
    Project & Observation & Configuration & Central Frequency & Flux & Phase \\
    {} & Date & {} & (GHz)  & Calibrator & Calibrator \\
    \hline
  AT141$^{\rm a}$    &   1992 Nov 24 (1992.9)      &     A    &     8.44  &    3C48     &    J2007+4029   \\
  AF381$^{\rm a}$    &   2001 Apr 23 (2001.31)     &     B    &     15.0  &    3C286    &    J2015+371   \\
  AS831$^{\rm a}$    &   2006 May 18 (2006.38)     &     A    &     8.46  &    3C286    &    J2007+4029    \\
  14A-007$^{\rm b}$  &   2014 March 15 (2014.29)   &     A    &     6.0   &    3C286    &    J2007+4029   \\
  14A-007$^{\rm b}$  &   2014 March 15 (2014.29)   &     A    &     15.0  &    3C286    &    J2007+4029   \\
  14A-007$^{\rm b}$  &   2014 March 15 (2014.29)   &     A    &     22.0  &    3C286    &    J2007+4029   \\
  14A-007$^{\rm b}$  &   2014 March 15 (2014.29)   &     A    &     44.0  &    3C286    &    J2007+4029   \\
\hline
  \end{tabular}
    \begin{tablenotes}
      \small
      \item (a) Archive data.
      \item (b) Partially reported by \citet{carrasco2015}.
    \end{tablenotes}
  \end{threeparttable}
 \end{table*}

\section{Results}

\subsection{VLA}\label{VLA-sources}

The most recent data set (epoch 2014.29, see Table \ref{tbl-images}) provides radio continuum images of W75N~(B) with unprecedented sensitivity. Also, these observations enable us to perform a detailed study of different structures associated with the sources and their emission nature within several ranges of frequencies.

By combining data from all the observed bands  (epoch 2014.29) we obtain the hitherto deepest (rms $\simeq$ 8~$\mu$Jy/beam; beam = 0$\farcs$12$\times$0$\farcs$09, PA = --69$^{\circ}$) radio continuum image of this region, and a spectral index map (Fig. \ref{w75n}). All the previously known sources in the field (VLA~1, VLA~2, VLA~3, VLA~4, and Bc) are labeled in Fig.~\ref{w75n}a.

In Fig. \ref{bands} we show the images of the radio continuum sources at C, Ku, K, and Q bands, covering the entire bandwidth in each case (see table \ref{tbl-images} for the image parameters). Due to spatial filtering effects that worsen at increasing frequency for a given configuration, C and Ku bands are more sensitive to extended emission than K and Q bands, which tend to favor the detection of compact structures. Taking this into account, and given the wide frequency coverage of our observations, both extended and compact emission of the sources can be studied with high sensitivity. We note that all the sources are detected with a S/N ratio much higher than 5 from C to K band (Figs. \ref{bands}a,b,c). However, at the highest frequency (Q band in Fig. \ref{bands}d), only the northern radio sources are detected, with Bc and VLA~4 hardly distinguishable from the noise. The flux densities of VLA~1, VLA~3, VLA~4, and Bc at each band are listed in Table \ref{tbl-srcFlux} \citep[a detailed discussion of the parameters of VLA~2 is given in][]{carrasco2015}. 

In addition to all these sources, three new weak ($<$100~$\mu$Jy, see Table \ref{tbl-srcFlux}) compact radio continuum sources are detected in the images at Ku and K bands, as well as in the image obtained by combining all four bands. These new sources are located $\sim$8 arcsec northeast (VLA [NE]) and $\sim$6 arcsec southwest (VLA [SW]) from VLA~2, and $\sim$0.5 arscec northeast (Bd) from VLA~4. Figs. \ref{figualma} and \ref{new_sources} show the position and contour maps, respectively, of these weak radio sources). 

\subsection{ALMA}\label{alma-sources}

The ALMA continuum observations at 1.3~mm show four cores in a region of $\sim$14~arcsec (MM1, MM2, MM3, MM[N]; Fig. \ref{figualma}). Three of them (MM1, MM2, MM3) have been previously identified by \citet{minh2010} with the Submillimeter Array (SMA) at 217 and 347~GHz, with angular resolution similar to that in our ALMA images. The fourth millimeter core, MM[N], is located $\sim$9~arcsec north from MM1 and has not been previously reported. We did not detect with the VLA any radio continuum source toward MM[N], with an upper limit of $\sim$30~$\mu$Jy (4$\sigma$ in the combination of C+Ku+K+Q bands; Table \ref{tbl-images}). From Fig. \ref{figualma} we see that the massive protostars VLA~1, VLA~2, and VLA~3 are associated with the brightest millimeter core, MM1, although the limited angular resolution of the ALMA observations ($1$\farcs$73\times0$\farcs$86$,  PA = $-4^\circ$) and the north-south distribution of the sources (with a maximum angular separation among them of $\lesssim$1.5~arcsec), prevent us from distinguishing the contribution of the individual sources to the total dust emission of MM1. On the other hand, two of the newly identified VLA radio continuum sources, VLA[SW] and VLA[NE] (Section \ref{VLA-sources}; Fig. \ref{figualma} and Fig. \ref{new_sources}), are associated with MM2 and MM3, respectively, indicating they are embedded YSOs. However, the fact that Bc, VLA~4, and Bd are not clearly associated with any millimeter continuum core suggests that they are probably not YSOs (see Section \ref{bc_vla4} for the discussion, in particular, on the nature of the Bc and VLA~4 radio continuum sources). 

Because these millimeter continuum cores are found at the edge of the ALMA primary beam and they were only observed at a single frequency band, we are not able to derive with accuracy their physical parameters with the present data. For those estimates we refer to \citet{minh2010}.

In Fig. \ref{figualma} we also show the images of the integrated intensity and velocity field (first order momentum) of the molecular lines CH$_3$OH [5(1,4)--4(2,2); rest frequency 216.94552 GHz], SO$_2$ [22(2,20)-22(1,21); 216.64330 GHz], and SiO [v=0 (5-4); 217.10498 GHz] as observed with the ALMA archive data. A main molecular core centered on VLA~1-VLA~2-VLA~3 is detected through the CH$_3$OH and SO$_2$ lines. This molecular core, of $\sim$4~arcsec ($\sim$5200~au) size, exhibits a velocity gradient of $\sim$5~km/s along the northwest--southeast direction, which is fully consistent with the velocity gradient reported by \citet{minh2010} in H$_2$CO with the SMA  ($\sim$1~km/s~arcsec$^{-1}$). Given their angular resolution, these ALMA observations \citep[and the SMA observations;][]{minh2010} cannot resolve the structure and motions of the dust and molecular gas around each of the individual sources VLA~1, VLA2, and VLA~3. 

On the other hand, the SiO emission shows an irregular distribution covering a broader velocity range (V$_{\rm LSR}$ $\approx$ 10--35~km/s) than CH$_3$OH and SO$_2$ (Fig. \ref{figualma}), with the highest velocity emission (V$_{\rm LSR}$ $\approx$ 30--35 km/s) located $\sim$6~arcsec northeast from VLA~1-VLA~2-VLA~3. This high-velocity SiO emission could be tracing outflow motions driven by any of the central VLA sources. ALMA observations with higher angular resolution are clearly necessary to identify, isolate, and study the expected different dust and molecular gas components around the individual massive protostars.


  \begin{table*}  \caption{Integrated flux densities of the radio sources detected in epoch 2014.29.}
   \center
 \label{tbl-srcFlux}
  \begin{threeparttable}
  \begin{tabular}{lccccc}
    \hline
 Source  &  S$_{\rm C}$ &  S$_{\rm Ku}$&  S$_{\rm K}$&  S$_{\rm Q}$ &  S$_{\rm CKuKQ}$\\
 {}     &  [mJy]  &  [mJy]  &  [mJy] &    [mJy] &  [mJy] \\
  \hline
VLA~1   &   3.8 $\pm$ 0.3   &   5.5 $\pm$ 0.3    &   5.8 $\pm$ 0.1     &  5.2 $\pm$  0.3    &  6.0 $\pm$ 0.2     \\
VLA~3   &   1.6 $\pm$ 0.1   &   5.2 $\pm$ 0.2    &   7.7 $\pm$ 0.1     &  15.7 $\pm$  0.8   &  9.06 $\pm$ 0.07   \\
VLA~4   &   0.67 $\pm$ 0.05 &   0.76 $\pm$ 0.05  &   0.83 $\pm$ 0.05   &  0.53  $\pm$ 0.08  &  0.69 $\pm$ 0.04   \\
Bc[E]   &   1.9 $\pm$ 0.2   &   1.8 $\pm$ 0.3    &   1.9 $\pm$ 0.3     &  $<$ 0.080         &  1.5 $\pm$ 0.2     \\
Bc[W]   &   1.7 $\pm$ 0.2   &   1.5 $\pm$ 0.2    &   1.4 $\pm$ 0.3     &  $<$ 0.080         &  1.4 $\pm$ 0.3     \\
Bd      &   $<$ 0.1         &   0.02$\pm$ 0.01    &   0.09 $\pm$ 0.02   &  $<$ 0.080         &  0.06 $\pm$ 0.01     \\
VLA[SW] &   $<$ 0.1         &   0.09$\pm$ 0.01    &   0.11 $\pm$ 0.02   &  $<$ 0.080         &  0.09 $\pm$ 0.01     \\
VLA[NE] &   $<$ 0.1         &   $<$ 0.052        &   0.03 $\pm$ 0.03   &  $<$ 0.080         &  0.041 $\pm$ 0.008   \\
\hline
  \end{tabular}
  \begin{tablenotes}
      \small
      \item Flux density upper limits correspond to 4$\sigma$. Errors are computed as the quadratic sum of both calibration and fitting error, except in the case of the last column, S$_{\rm CKuKQ}$, which correspond to fitting errors. See Table \ref{tbl-images} for image details.
    \end{tablenotes}
    \end{threeparttable}
 \end{table*}

\section{Nature of the individual radio sources}\label{sec-discussion}

\subsection{VLA 1}\label{sec-vla1}

\begin{figure*}
    \centering
    \includegraphics[width=\textwidth]{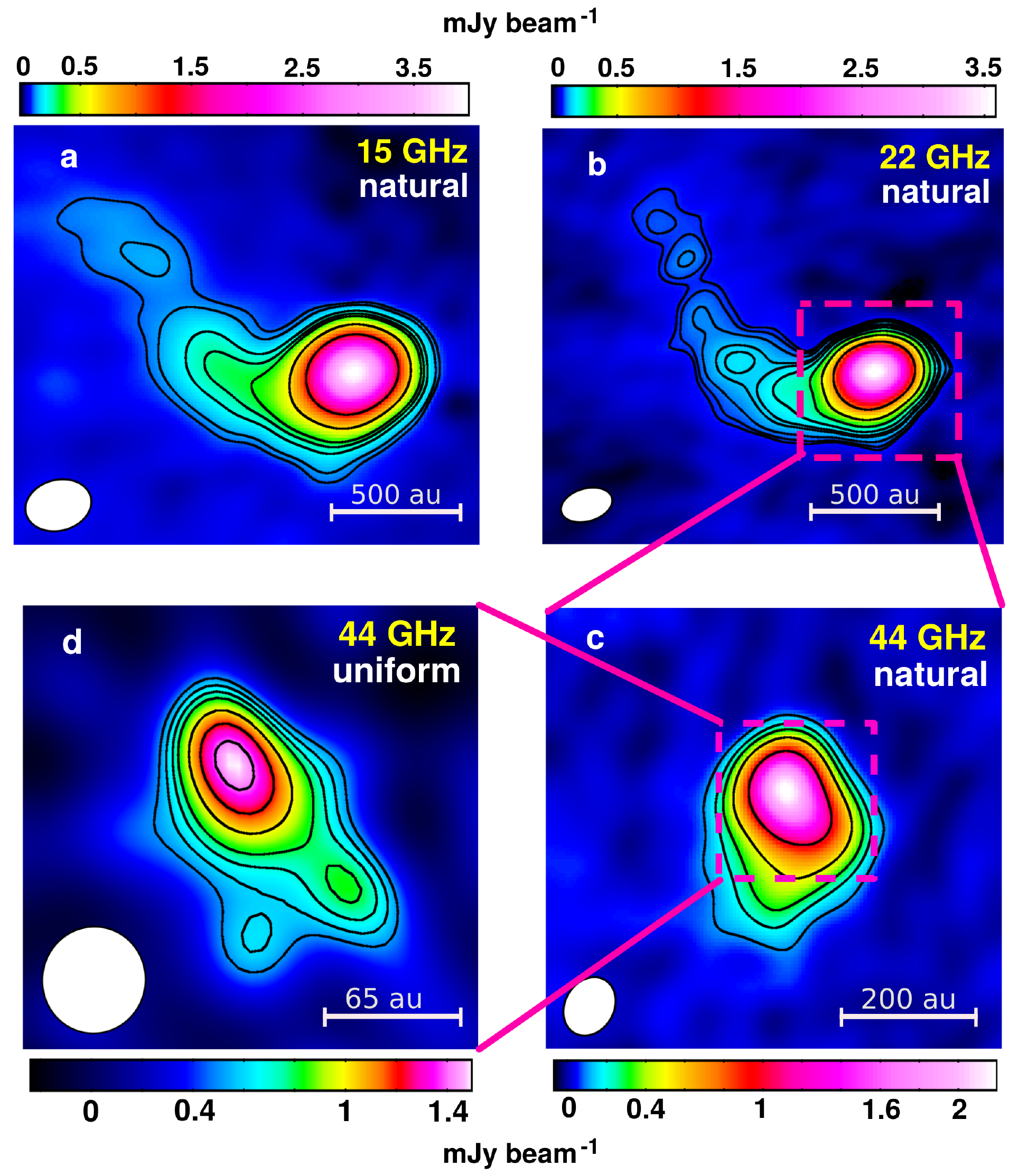}
    \caption{VLA~1 continuum emission (epoch 2014.29). (a) Natural-weighted image at Ku band (central frequency 15~GHz); contours are --3, 6, 8, 15, 20, 25, 50, and 100 $\times$ 10~$\mu$Jy/beam, the rms of the map. The synthesized beam is 0$\farcs$21 $\times$ 0$\farcs$16, PA = -75$^{\circ}$. (b) Natural-weighted image at K band (central frequency 22~GHz); contours are --3, 6, 7, 9, 13, 15, 20, 30, 50, 100 $\times$ the rms = 10~$\mu$Jy/beam. The synthesized beam is 0$\farcs$16 $\times$ 0$\farcs$10, PA = -74$^{\circ}$. The magenta dotted rectangle indicates the zoomed-in region shown in panel (c). (c) Natural-weighted image at Q band (central frequency 44~GHz); contours are --3, 5, 10, 15, 25, 50 $\times$ the rms = 20~$\mu$Jy/beam. The synthesized beam is 0$\farcs$07 $\times$ 0$\farcs$05, PA = -29$^{\circ}$. The magenta dotted rectangle indicates the zoomed-in region shown in panel (d). (d) Uniform-weighted image at Q band; contours are -3, 5, 6, 7, 8, 10, 12, 14 $\times$ the rms = 100~$\mu$Jy/beam. The synthesized beam is 0$\farcs$038 $\times$ 0$\farcs$037, PA = --9$^{\circ}$. The physical scale of 65 au corresponds to 0.05~arcsec.}\label{VLA1_bandas}

\end{figure*}

\begin{figure}
    \centering
    \includegraphics[width=\columnwidth]{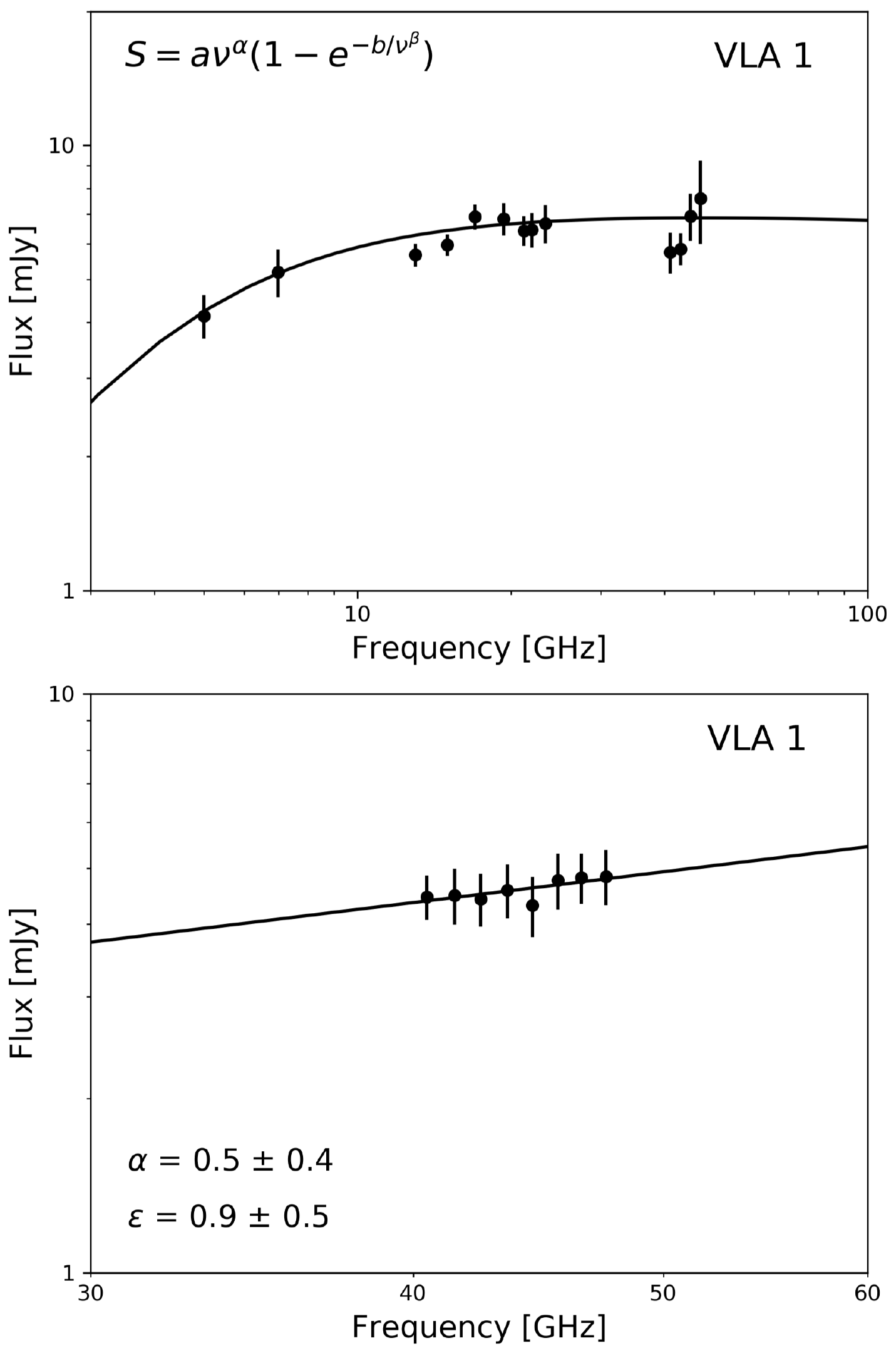}
    \caption{VLA 1 spectral energy distribution (epoch 2014.29). Top panel: the SED is computed in the whole range of observed frequencies. Flux densities are obtained from Gaussian fits to 2 GHz bandwidth images within a circular region of 1.25~arcsec diameter enclosing the source. We use uniform weighting at C band and natural weighting at Ku, K, and Q bands.
    A fit to the measured flux densities is also shown (S$_{\nu}$ = a$\nu^{\alpha}$ [1$-$e$^{-b/\nu^{\beta}}$], with a = 0.62, $\alpha$ = 1.38, b = 13.29, and $\beta$ = 1.42, with S$_{\nu}$ in mJy and $\nu$ in GHz). This fitted ad hoc function is only for description purposes of the observed SED (see Section \ref{sec-vla1}).
    Bottom panel: Spectrum at Q band. Flux densities are obtained from Gaussian fits to 1~GHz bandwidth images within a circular region of 0.16~arcsec diameter. We use Briggs weighting (robust 0). The solid line is a linear least-squares fit to the log data, from which we derive a spectral index $\alpha=+0.5\pm0.4$ that correspond to $\epsilon=+0.9\pm0.5$, consistent with a thermal radio jet (see Section \ref{sec-vla1}). All data points are shown with measurement errors, considering both fitting and calibration uncertainties. Note that these panels trace different components in the source: while the top panel corresponds to the HCHII region, the bottom one traces the compact jet.}
\label{VLA1_sed}
\end{figure}

\begin{figure}
    \centering
    \includegraphics[width=\columnwidth]{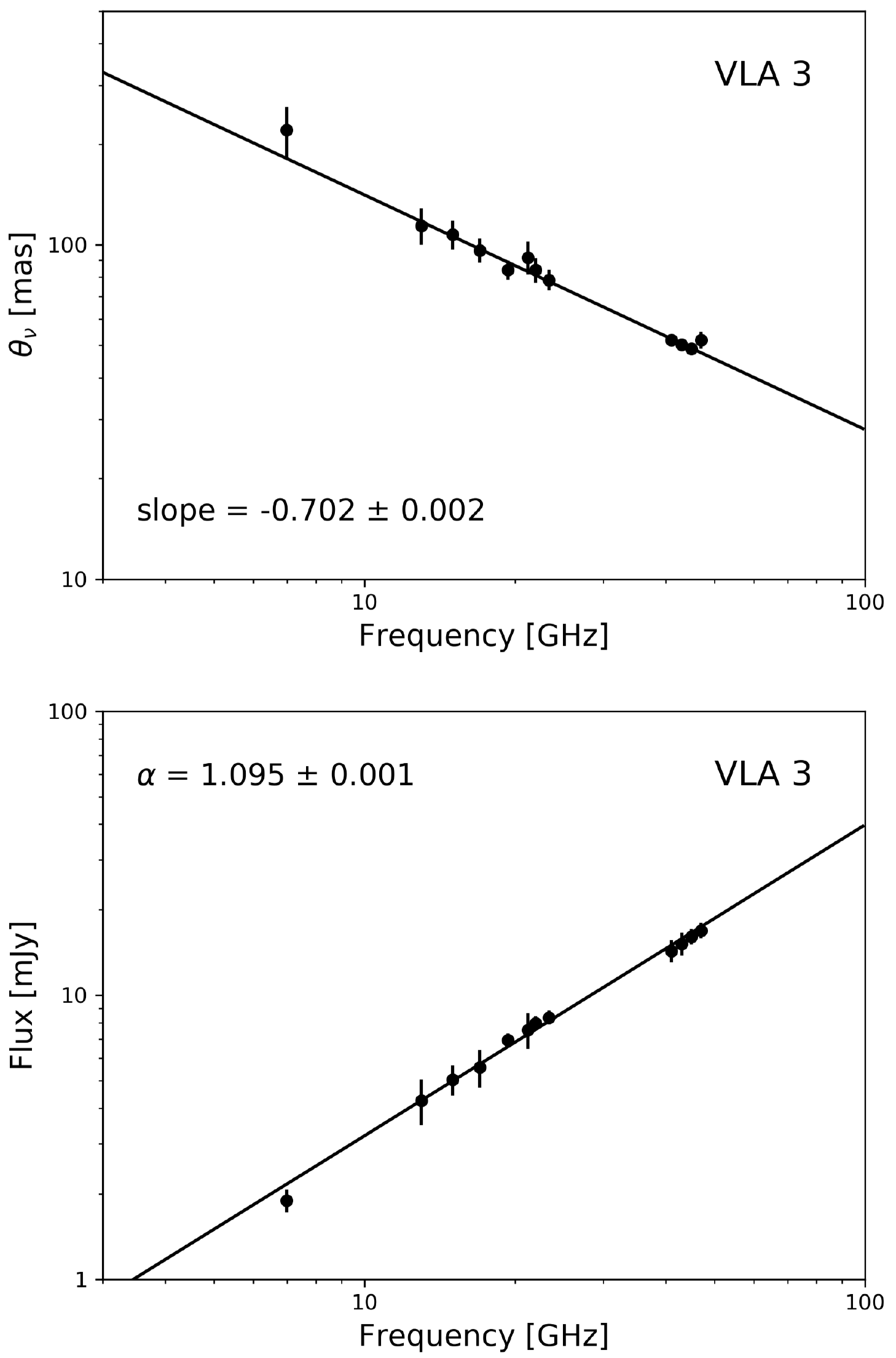}
    \caption{Dependence of the angular size of the jet with frequency (top panel) and spectral energy distribution (bottom panel) of VLA~3. Flux densities and semi-major axes $\theta_{\nu}$, are obtained from Gaussian fits to the brightness profile of VLA 3, in 2 GHz bandwidth images. To measure flux densities we use uniform-weighted images (C band) and natural-weighted images (Ku, K, and Q bands), while angular sizes were measured in uniform-weighted images (C, Ku, and K bands) and Briggs-weighted (robust -1) images (Q band). Solid lines are linear least-squares fits to the log data. Angular size error bars correspond to fitting errors, while flux error bars consider both fitting and calibration errors.}\label{fig-vla3}
\end{figure}

 It can be seen from Figs.~\ref{w75n} and \ref{VLA1_bandas} that VLA~1 exhibits a tail-shaped extended component. This is the first time that this particular structure is observed in VLA~1, due to the high sensitivity of our images. In Fig.~\ref{w75n}b we show the spatial distribution of spectral indices ($\alpha$, defined as S$_{\nu}$ $\propto\nu^{\alpha}$) along the source, covering the whole range of frequencies. It can be noticed that the central region of VLA~1 presents positive spectral indices ($\sim+$0.5, measured from the spectral index map, at the continuum emission peak), while the tail-shaped structure is dominated by a flat spectrum with $\alpha\simeq 0$.
 
By studying the emission at different bands, we can see that the morphology of the VLA 1 source seems to be composed by an extended and a compact components (see Figure \ref{VLA1_bandas}). At the lowest frequencies (15 and 22 GHz), the higher optical depth and lower angular resolutions emphasize the extended component. In Figures \ref{VLA1_bandas}a and \ref{VLA1_bandas}b, we can clearly see a "cometary" tail curved in the northeast direction. As we go to higher frequencies (44 GHz), the higher angular resolution and lower optical depth allow us to filter out most of the low-brightness extended emission, and the most compact higher brightness emission is resolved into an elongated source in the NE-SW direction (PA = $+42^{\circ}\pm5^{\circ}$; see Figures \ref{VLA1_bandas}c and \ref{VLA1_bandas}d).
 
To study the emission nature of both, the extended and the compact components of VLA~1, we compute the spectral energy distribution (SED) of the source in the whole range of observed frequencies with low angular resolution, and the SED at Q band (where most of the extended emission is filtered) with high angular resolution. In Fig.~\ref{VLA1_sed} (top panel) we show the SED over the entire range of frequencies, obtained by measuring flux densities in images with 2~GHz bandwidth, using uniform weighting for C band and natural weighting for Ku, K, and Q bands. All images were convolved to 0$\farcs$37, corresponding to the lowest resolution in C band. Flux densities were determined by a Gaussian fit within a circular region of 1.25 arcsec diameter enclosing the source.  We note that these data were observed with the telescope array same configuration, and in this case, the largest scales in the images could be more heavily filtered at high frequencies, which could result in spuriously lower values of the spectral index. However, we limited the study of the low angular resolution SED to the core of the emission, which has a size of $\sim$400 milliarcsec (mas). Emission with this size is fully recovered at all bands. Just for description purposes, we have performed a fit to the observed flux densities from 4 to 47 GHz through an ad hoc function S$_{\nu}$ = a$\nu^{\alpha}$ [1$-$e$^{-b/\nu^{\beta}}$] (Fig.~\ref{VLA1_sed}, top panel). The fit gives a = 0.62, $\alpha$ = 1.38, b = 13.29, and $\beta$ = 1.42, with S$_{\nu}$ in mJy and $\nu$ in GHz. Within the uncertainties in the observations, this SED is consistent with an HII region thermal bremsstrahlung spectrum, opaque at low frequencies ($\lesssim$ 10~GHz) and optically thin at high frequencies ($\gtrsim$ 20 GHz).  The size of the extended emission, including the tail, is of the order of 1~arcsec, corresponding to an extension of $\sim$0.006~pc. This size is significantly smaller than 0.1~pc, suggesting it could be classified as a Hypercompact (HC) HII region, according to \citet{kurtz2005}. 
 
 We want to note, however, that we cannot rule out the possibility that some dust contribution from the extended emission in VLA~1 is present at Q-band (Figs. \ref{VLA1_bandas}a and \ref{VLA1_bandas}b). If this were true, the real spectral energy distribution of the ionized gas would be flatter at high frequencies than that shown in Fig.~\ref{VLA1_sed} (top panel). In any case, as stated in Section~\ref{alma-sources}, very high angular resolution observations with ALMA are clearly needed to individually resolve the dust content of each YSO in the region.
 
 In order to study VLA~1 with as less contribution as possible from the extended emission, we measured flux densities with very high angular resolution in eight R0-weighted (Briggs weighting using parameter robust $=$ 0) images of 1~GHz bandwidth within the Q band (see table \ref{tbl-images} for image details). All flux densities are measured within a circular region of 0.16~arcsec diameter enclosing the source. The resulting SED is shown at the bottom panel of Fig. \ref{VLA1_sed}. Applying a linear fit to this SED, we obtain a spectral index $\alpha= +0.5\pm0.4$, consistent with partially optically thick free-free emission from a thermal radio jet, as predicted by models given by \cite{reynolds1986}, and consistent with typical values measured for thermal radio jets \citep[e.g.,][]{anglada2018}. This is also in agreement with previous works by \citet{baart1986} and \citet{torrelles2003}, who detected the presence of a radio jet traced by the distribution of OH and H$_{2}$O masers, respectively, along the same direction as the radio continuum emission \citep[PA $\approx$ +43$^\circ$,][]{torrelles1997,torrelles2003}. According to Reynolds models, the distance to the driving source where the jet becomes optically thin varies with frequency as a power law. This distance is interpreted as the angular size of the semi-major axis of the jet, $\theta_{\nu}$. Therefore, $\theta_{\nu}\propto\nu^{-0.7/\epsilon}$, where the index $\epsilon$ is related to the spectral index $\alpha$ as $\alpha=1.3-0.7/\epsilon$ in the case of an isothermal jet with constant velocity and ionization fraction. Thus, from the SED in Q band we derive $\epsilon= +0.9\pm0.5$, suggesting a conical jet ($\epsilon=1$).
 
 All these results indicate that VLA~1 might consist of a thermal radio jet surrounded by an HCHII region. Assuming this is the case, we can estimate the mass-loss rate and some parameters of the HII region. In order to estimate the protostar mass-loss rate $\dot{M}$, we follow Equation (3) from \cite{beltran2001} based on the model from \cite{reynolds1986}:

\begin{multline}\label{massloss}
    \frac{\dot{M}}{\rm 10^{-6}~M_{\odot}~yr^{-1}}=0.108\left[\frac{(2-\alpha)(\alpha+0.1)}{1.3-\alpha}\right]^{0.75}\\
    \times~\left[\left(\frac{S_{\nu}}{\rm mJy}\right)\left(\frac{\nu}{\rm 10~GHz}\right)^{-\alpha}\right]^{0.75}\left(\frac{d}{\rm kpc}\right)^{1.5}\left(\frac{\nu_{t}}{\rm 10~GHz}\right)^{0.75\alpha-0.45}\\
    \times~\left(\frac{\theta_{o}}{\rm rad}\right)^{0.75}\left(\frac{T_e}{\rm 10^{4}~K}\right)^{-0.075}(\sin{i})^{-0.25}\left(\frac{V_{\rm jet}}{200~{\rm km~s^{-1}}}\right)\left(\frac{1}{x_{o}}\right),
\end{multline}

\noindent where $S_{\nu}$ is the flux density at the frequency $\nu$, $\alpha$ the spectral index, $\theta_{o}=0.88$~rad the jet injection opening angle, estimated as $\theta_{o}=2\arctan(\theta_{\rm min}/\theta_{\rm maj}$), where $\theta_{\rm min}$ and $\theta_{\rm maj}$ are the deconvolved minor and major axes of the Gaussian fit to the source respectively. T$_e$ = $10^{4}$~K is the electron temperature, $x_{o}$ the ionization fraction, V$_{\rm jet}$ the jet velocity, $i$ the jet inclination angle, $\nu_{\rm t}$ the turn-over frequency, and d = 1.3~kpc the distance to the region W75N~(B). The values of $\nu$ and $S_{\nu}$ correspond to the Q-band image, being $\nu$ = 44~GHz the central frequency of the band and $S_{\nu}$ = 15.7$\pm$0.1~mJy. The spectral index $\alpha$ = +0.5$\pm$0.4 is derived from a linear fit to the SED (see Fig. \ref{VLA1_sed}, bottom panel), from which we can also infer that the jet emission is partially optically thick up to $\nu$= 47~GHz. Thus, we take this value as a lower limit to the turn-over frequency $\nu_{\rm t}$ (above which the entire jet becomes optically thin). Since we cannot specify the jet inclination angle $i$, we adopt $i=$45$^{\circ}$, as variations from 45$^{\circ}$ to 90$^{\circ}$ only change the mass-loss estimate by less than 10\%. 
Moreover, both the ionization fraction and the jet velocity are unknown. Typical values for V$_{\rm jet}$ range from 100 to 1000 km~s$^{-1}$, while the ionization fraction is usually assumed to be 10\% \citep[e.g.,][]{anglada2018} for low-mass protostars. However, this value is very uncertain, and could probably be higher for high mass protostars. According to this, we estimate lower and upper limits for the mass loss rate of $\sim$3.5$\times$10$^{-7}$~M$_{\odot}$~yr$^{-1}$ (assuming $x_{o}$ = 1 and V$_{\rm jet}$ = 100~km~s$^{-1}$) and $\sim$3.5$\times$10$^{-5}$~M$_{\odot}$~yr$^{-1}$ (assuming $x_{o}$ = 0.1 and V$_{\rm jet}$ = 1000~km~s$^{-1}$), respectively. 
 
 On the other hand, we can obtain some estimates of the physical parameters of the HII region using data from C to K bands. Derived parameters are presented in Table \ref{tbl-vla1ndot}. Within the Rayleigh-Jeans regime, the brightness temperature T$_{\rm B}$ can be written in terms of the flux density $S_{\rm \nu}$ at frequency $\nu$, and the solid angle subtended by the source $\Omega_{\rm S}$, as $T_{\rm B}=\frac{S_{\rm \nu}c^{2}}{2k\nu^{2}\Omega_{\rm S}}$. The solid angle $\Omega_{S}$ of the elliptical Gaussian fitted to the source brightness profile is calculated as $\rm{(\pi/4\ln2)\times FWHM_{maj}\times FWHM_{min}}$. At the central frequency of each band, T$_{\rm B}$ is always small compared with the electronic temperature T$_{\rm e}$, assumed to be of the order of $10^{4}$~K for an HII region (since this is the temperature at which hydrogen ionizes). Knowing the brightness temperature, the optical depth $\tau_{\nu}$ can be calculated from $\mathrm{T_{B} = T_{e}(1-e^{-\tau_{\nu}})}$. As the HII region is more optically thin at the K band, we choose this band to estimate some parameters. Assuming the ionization number equals the recombination number, we can estimate the ionizing photon rate $\dot{\rm N}$, i.e. the number of ionizing photons $\lambda<912~\textup{\r{A}}$ per unit of time, necessary to account for the emission observed at K band:

 \begin{equation}\label{n_dot}
    \dot{\rm N}=\frac{4}{3}\pi\left(\frac{L}{2}\right)^{3}n_{e}n_{p}\eta,
\end{equation}

\noindent where $\eta=3\times10^{-13}$~cm$^3$s$^{-1}$ the "case B" recombination coefficient (i.e., the number of recombinations per unit time, volume, and electron and ion density) to levels $\geq$2 for T$_e$ $\approx$10$^4$~K; $n_{e}$ and $n_{p}$ are, respectively, the number density of electrons and protons (assumed to be the same), and $L$ is the characteristic size of the region (estimated as the geometric mean of the source major and minor axes). Assuming a homogeneous, spherical HII region of depth L, $n_{e}$ can be expressed in terms of the emission measure (EM) and the geometrical depth as $n_{e}=(\rm EM/L)^{1/2}$ (with ${\rm EM =  \int_{0}^{L} n_{e}n_{p}~dl}$ which, in the case of a homogeneous ionized hydrogen medium of depth L approximates to $EM\simeq n_e^2L$). In turn, EM can be derived from the expression of the optical depth as $\rm EM[{\rm cm^{-6}~pc}]= 12.2 \tau_{\nu}[T_{e}/{\rm K}]^{1.35}[\nu/{\rm GHz}]^{2.1}$. Thus, computing ${\rm S}_{\nu}$, $\Omega$, T$_B$, $\tau_{\nu}$, L, ${\rm n}_{e}$, and EM, at the central frequency of band K (Table \ref{tbl-vla1ndot}), we finally obtain an ionizing photon rate $\dot{\rm N}$ $\approx$ 6$\times10^{44}$ photons s$^{-1}$. This value is much lower than typical estimations for O-B stars ($\approx$10$^{48}$ photons~s$^{-1}$). Moreover, characteristic values of emission measure and electron number density for HCHII regions are EM $\gtrsim10^{10}$~pc~cm$^{-6}$ and $n_{e}\gtrsim10^6$~cm$^{-3}$, respectively \citep{kurtz2005}. In the case of VLA~1, the electron number density we obtain is of the same order of typical values for HCHII regions, while the emission measure is about two orders of magnitude lower (see Table \ref{tbl-vla1ndot}). 

Mass-loss rates in the range of 
$10^{-10}~M_\odot$~yr$^{-1}$ (for low-mass YSOs) to 
$10^{-5}~M_\odot$~yr$^{-1}$ (for high-mass YSOs) have been determined in the literature \citep[see][and references therein]{anglada2018}. Thus, our estimates for the mass-loss rate of VLA~1, together with the physical parameters we obtain for the HII region, lead us to conclude that VLA~1 might be a massive protostar driving a thermal radio jet, which seems to be at the very beginning of the photoionization stage. The presence of a radio jet coexisting with an UCHII has been also reported in the massive YSOs G35.20--074N \citep{beltran2016} and G345.4938+01.4677 \citep{guzman2016}.
 
 \begin{table}     \caption{VLA~1 PARAMETERS FROM K-BAND EMISSION}
 \label{tbl-vla1ndot} 
  \begin{threeparttable}
  \begin{tabular}{lll}
    \hline
   Parameter  &  Value  &  Description\\
    \hline
$\nu$             &   22$\times$10$^{9}$~Hz                      &   band central frequency  \\
$S_{\nu}$               &   (5.6 $\pm$ 0.2)~mJy                          &   flux density            \\
$\rm{FWHM_{min}}$    &   (74 $\pm$ 3)~mas = (3.6 $\pm$ 0.2)$\times$10$^{-7}$~rad             &   source minor axis  \\
$\rm{FWHM_{maj}}$    &   (100 $\pm$ 2)~mas = (4.8 $\pm$ 0.1)$\times$10$^{-7}$~rad             &   source major axis \\ 
L                 &   $\sim$90~mas $\simeq$110~au &   source characteristic size  \\
$\Omega_{\nu}$          &   (2.0 $\pm$ 0.2) $\times$10$^{-13}$~sr       &  subtended solid angle  \\
T$_{\rm B}$       &   (1890 $\pm$ 10)~K                             &  brightness temperature  \\
$\tau_{\nu}$            &   $\sim$0.2                          &  optical depth  \\
EM                &   $\sim$4$\times10^{8}$~cm$^{-6}$pc  &  emission measure  \\
$n_e$             &   $\sim$$10^{6}$~cm$^{-3}$       &  electron number density  \\
$\dot{\rm N}$     &   $\sim$6$\times10^{44}$~photons s$^{-1}$  &  ionizing photon rate      \\          
\hline
  \end{tabular}
  \begin{tablenotes}
      \small
      \item The flux density $S$, $\rm{r_{min}}$, $\rm{r_{maj}}$, and the characteristic size L correspond to values derived from a by-dimensional Gaussian fit to VLA~1 at K-band (robust 0 weighting).
    \end{tablenotes}
    \end{threeparttable}
 \end{table}

\subsection{VLA 3}\label{vla3}

VLA~3 was previously proposed to be a partially optically thick compact HII region \citep[e.g.,][]{torrelles1997,shepherd2004}, but it was later suggested to be a thermal radio jet with a spectral index $\alpha_{3.6-2{\rm cm}}=+0.6\pm0.1$ \citep{carrasco2010}.  In our data (Figs. \ref{w75n} and \ref{bands}), VLA~3 appears as an elongated source, with its major axis oriented in the northwest-southeast direction at all wavelengths, with a position angle PA = -17$^{\circ}\pm$2$^{\circ}$ (Fig. \ref{w75n}). The ionized emission is characterized by a spectral index $\alpha$ $\simeq$ $+$1 in the central region of VLA~3 (Fig. \ref{w75n}b), consistent with partially optically thick free-free emission. Its elongated shape and its spectral index suggest that VLA~3 could be a thermal radio jet. Thus, according to theoretical models given by \citet{reynolds1986}, we expect to find that both the SED and the jet angular size depend on the observed frequency as power laws (see Section \ref{sec-vla1}). Therefore,  we measured the flux densities and angular sizes in several images of 2~GHz bandwidth each, covering a frequency range from 7 to 48 GHz. Flux densities are measured within a circular region of 0.5~arcsec diameter in each image, and the angular sizes correspond to the deconvolved major axis of the bi-dimensional Gaussian fit to the emission. Sizes vary in the range 30-200~mas, and therefore, we can rule out that a significant amount of extended emission is filtered out at the highest frequencies. In Fig. \ref{fig-vla3} we can see that both the angular-size $\theta_{\nu}$ (top panel) and the SED (bottom panel) do vary as power laws of the frequency. In the ideal case of a conical thermal jet, with constant velocity, temperature, and ionization fraction, values of +0.6 and $-$0.7 are expected for the spectral index and the slope of the size vs frequency, respectively \citep{reynolds1986}. In our case, the behavior of the size with frequency is consistent with a conical jet, while the slightly larger value of the spectral index would indicate some deviation from the ideal physical conditions.

These results support that VLA~3 is associated with a thermal radio jet as previously proposed by \citet{carrasco2010}. Therefore, in order to estimate the protostar mass-loss rate $\dot{M}$ we follow Equation \ref{massloss}. In this case $\nu$, $S_{\nu}$, and $\alpha$ correspond to the combined image (C+Ku+K+Q bands), being $\nu$ = 26~GHz the central frequency of the band, $S_{\nu}$ = 9.06$\pm$0.07~mJy, and $\alpha$ = $+$1.27 (computed from the spectral index map at the continuum emission peak). From the SED (bottom panel of Fig. \ref{fig-vla3}) we can see that the emission is partially optically thick up to $\nu$= 47~GHz, thus, we take this value as a lower limit to the turn-over frequency $\nu_{\rm t}$. As in the case of VLA~1, we also adopt a jet inclination angle $i=$45$^{\circ}$, and estimate lower and upper limits for the mass loss rate considering different approximations for the ionization fraction and the jet velocity, i.e., $\sim$4$\times$10$^{-6}$~M$_{\odot}$~yr$^{-1}$ (assuming $x_{o}$ = 1 and V$_{\rm jet}$ = 100~km~s$^{-1}$) and $\sim$4$\times$10$^{-4}$~M$_{\odot}$~yr$^{-1}$ (assuming $x_{o}$ = 0.1 and V$_{\rm jet}$ = 1000~km~s$^{-1}$), respectively. Such mass-loss rates are significantly higher than those estimated in low- and intermediate-mass YSOs \citep[e.g.,][]{beltran2001, anglada2018}, but similar to the values obtained in high-mass YSOs  \citep[e.g.,][]{lfr1994, guzman2012, anez2020}, supporting that VLA~3 is excited by a massive protostar.

\subsection{Bc and VLA 4}\label{bc_vla4}

 In Fig. \ref{w75n}c we show the radio image with the highest resolution and sensitivity to date of Bc and VLA~4. This allows us to resolve their structure, and study the nature of their emission through the spectral index map. The source Bc is clearly resolved into two components (labeled as Bc~[E] and Bc~[W] in Fig. \ref{w75n}). We note that Bc~[W]--Bc~[E] form an elongated structure, with its minor axis aligned with the VLA~3 jet direction as we would expect to observe in a bow-shock produced by the impact of a supersonic jet with the environment gas \citep[e.g.,][]{tafalla2017,caste2018}. 
 This supports the scenario proposed by \citet{carrasco2010} who interpreted Bc as an obscured radio Herbig-Haro (HH) object, possibly excited by the VLA~3 jet. A flattened structure similar to that of Bc is also seen in the frontal region of the shock of the obscured HH~80N object \citep[][]{ark2019}.

  \begin{figure}
    \centering
    \includegraphics[width=\columnwidth]{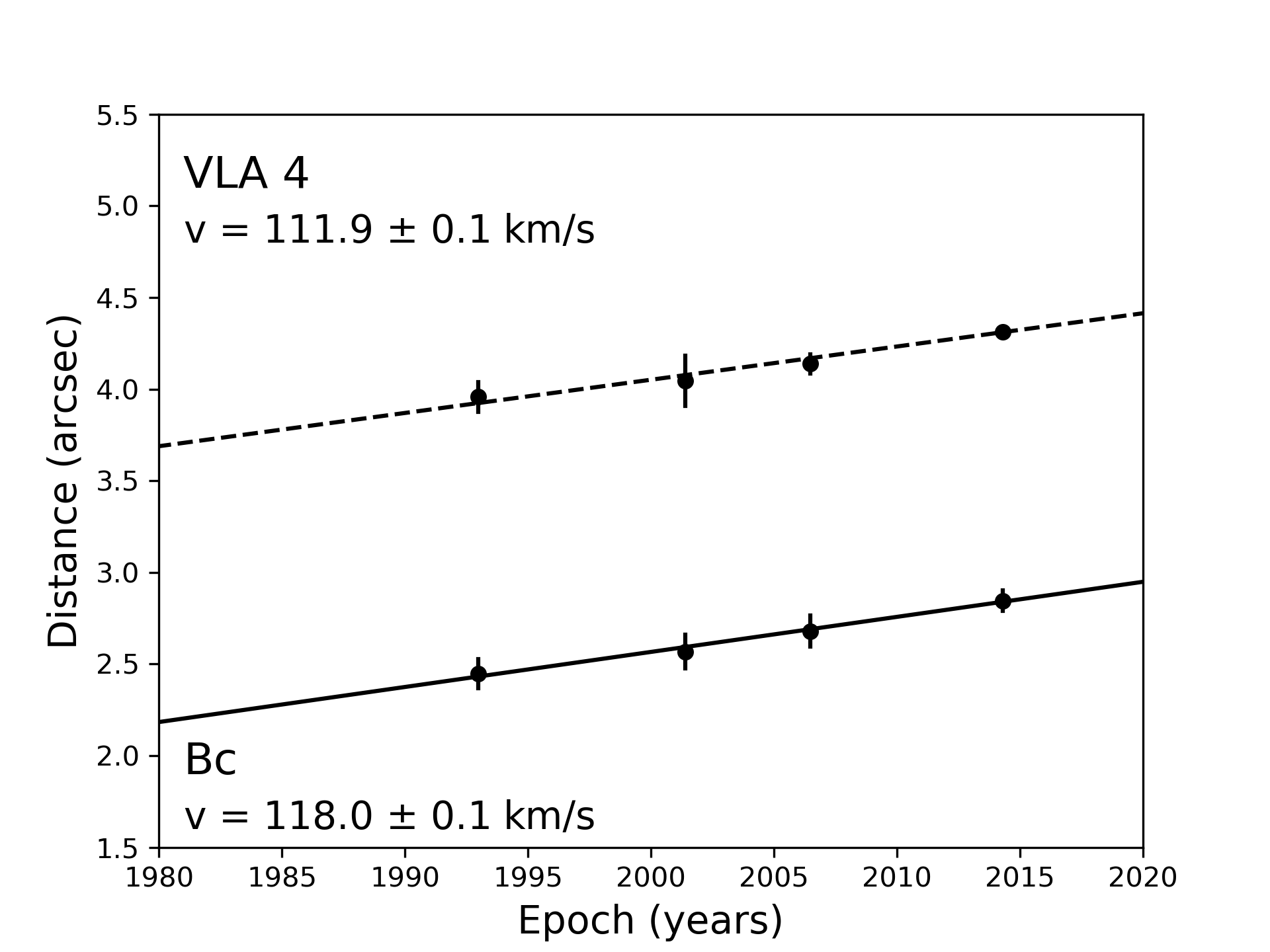}
    \caption{Proper motion diagrams for the Bc and VLA~4 sources. Positions are computed as the distances to the averaged coordinates of the system VLA~1-VLA~2-VLA~3 in four epochs (1992.98, 2001.40, 2006.47, 2014.29): RA(J2000) = 20h 38m 36.47s, DEC(J2000)= 42$^{\circ}$ 37$^{\prime}$ 34.15$^{\prime\prime}$. The solid lines are least-square fits to the data. Velocities on the plane of sky are derived by assuming a distance to the region of 1.3 kpc \citep{rygl2012}. Bc and VLA~4 are moving away from the system with PAs of $\sim$-20$^{\circ}$ and $\sim$-10$^{\circ}$, respectively.}\label{pm}
\end{figure}

 \citet{carrasco2010} studied the kinematics of these sources by computing proper motions relative to VLA~3 in three epochs (1992.90, 1998.23, and 2006.38) spanning 13.48 years. Adopting a distance to the region of 2~kpc \citep{dickel1969}, they derived for Bc a velocity of 220$\pm$70~km~s$^{-1}$ moving on the plane of the sky, and toward the south, approximately along the major axis of the VLA~3 radio jet. Regarding VLA~4, they suggested it could be either an independent star or shock-excited gas produced by a previous ejection from VLA~3. However, even though they noticed a small displacement of VLA~4 to the south with respect to VLA~3 between 2000 and 2006, they were not able to distinguish between these two scenarios. Regarding this, we measured the proper motion of both sources along a time span of 22 years, from 1992 to 2014. Positions of the sources in 2014 were measured in the K band image, since it is the highest angular resolution image where Bc and VLA~4 are detected (see Fig. \ref{bands}). We compute the proper motions relative to the average position of the system VLA~1-VLA~2-VLA~3 assuming it is stationary, instead of VLA~3 only. In this way, we reduce possible errors associated to variations in the shape of VLA~3 due to observational effects (e.g. differences in the beam size at each epoch) or real changes (e.g., new ejections from VLA 3). Although Bc shows substructure, we measure the displacement of the two-component complex. 
 Fig. \ref{pm} shows that both Bc and VLA~4 are moving away from the system in the southeast direction (PA $\simeq$ -20$^{\circ}$ and -10$^{\circ}$, individually, see Fig. \ref{pm}) with velocities on the plane of the sky of 1.9$\times$10$^{-2}$~arcsec~yr$^{-1}$ ($\sim$118~km~s$^{-1}$), and 1.8$\times$10$^{-2}$~arcsec~yr$^{-1}$ ($\sim$112~km~s$^{-1}$), respectively \citep[assuming the updated distance to the region of 1.3 kpc;][]{rygl2012}. 
 The velocity of Bc differs from the value given by \citet{carrasco2010}, but this is mostly due to the difference in the adopted distance to the region (considering a distance of 2~kpc we obtain velocities of 170 and 180 km~s$^{-1}$ for VLA~4 and Bc, respectively, closer to the results reported by these authors). Thus, the shape and proper motions of Bc and VLA~4 are consistent with both sources tracing shock-excited gas (obscured HH objects). Moreover, from the spectral index map (Fig. \ref{w75n}c) we see that Bc and VLA~4 are dominated by flat spectral indices ($\alpha\sim$ 0), as it would be expected for optically thin free-free emission produced by shock-ionized material. Thus, flat spectral indices, together with the shape of the sources, their proper motions, and propagation direction constitute solid evidence supporting the shock scenario.

Among the sources VLA~1, VLA~2, and VLA~3, VLA~1 and VLA~2 have associated outflows along the northeast--southwest direction \citep[e.g.,][]{torrelles1997,torrelles2003,carrasco2015}. Therefore, VLA~3 is the only source elongated in the northwest--southeast direction (PA $\simeq$ -17$^{\circ}$), consistent with the direction of the proper motions of Bc and VLA~4 (PA $\simeq$ -20$^{\circ}$ and -10$^{\circ}$). This, along with the results found in section \ref{vla3}, suggests that VLA~3 is the driving source of Bc and VLA~4. In addition, the fact that Bc and VLA~4 are not associated with any of the detected millimeter cores (see Section \ref{alma-sources}; Fig. \ref{figualma}) suggests that they are not protostars, further supporting our shock-excited gas interpretation for these sources. This kind of obscured HH objects, exhibiting proper motions higher than 100~km~s$^{-1}$, have also been observed in radio continuum in other intermediate- and 
high-mass star-forming regions:  e.g., Serpens \citep[][]{curiel1993,ark2016}, GGD~27 \citep[][]{marti1995,marti1998,masque2015,ark2019}, Cepheus~A \citep[][]{curiel2006}.
  
\section{Conclusions}

We presented an analysis of high-sensitivity, high-resolution multi-frequency VLA observations of the massive star-forming region W75N (B), together with complementary studies performed with ALMA and VLA archive data. Our study leads us to the following conclusions:

\begin{itemize}
    \item VLA~1 is detected at all the observed frequencies (4-48 GHz). Its SED over the entire range of frequencies is consistent with thermal free-free emission from an HCHII region ($\lesssim$1 arcsec, $\lesssim$1300~au), while the high angular resolution spectrum of the most compact component at high frequencies (40-48~GHz) is compatible with a thermal radio jet at scales of $\approx$0.1~arcsec ($\approx$130~au), with a spectral index $\alpha\approx+0.5$ (S$_{\nu}$ $\propto\nu^{\alpha}$). This suggests that VLA~1 is driving a thermal radio jet, and it is likely at the early stage of the photoionization.
 
    \item VLA~3 shows an elongated structure at scales of few tenths of arcsec (few hundred of au), with its major axis oriented in the northwest--southeast direction (PA $\approx$ -17$^{\circ}$). Both the SED and the size dependence with frequency indicates that this source is also driving a thermal radio jet. 
    
    \item We computed proper motions of the radio continuum sources Bc and VLA~4 in a time interval of 22 years. We found both sources are moving away toward the south, in a similar direction as the VLA~3 thermal radio jet, with velocities of $\approx$112--118~km~s$^{-1}$ ($\approx$1.8--1.9$\times$10$^{-2}$~arcsec~yr$^{-1}$). 
    From the SED analysis we found these sources are dominated by flat spectral indices, as it is expected for optically thin free-free emission produced by shock-ionized material. These results support the scenario in which Bc and VLA~4 are obscured HH objects tracing shocks of the jet driven by VLA~3.

    \item Four 1.3 mm continuum cores are observed with ALMA (MM1, MM2, MM3, and MM[N]) in a region of $\sim$14 arcsec. Three of these millimeter cores, MM1, MM2, and MM3, had previously been identified with the SMA interferometer, while MM[N] had not been previously reported. VLA~1, VLA~2, and VLA~3 are associated with the brightest core MM1. 
    Bc and VLA~4 are not associated with any of the millimeter continuum cores, supporting they are not YSOs but shock-excited gas as concluded from our VLA observations.
    
    \item We have detected three new weak compact radio continuum sources (VLA[SW], VLA[NE], and Bd). Two of them, VLA[SW] and VLA[NE] are associated with the millimeter cores MM2 and MM3, respectively, suggesting they are embedded YSOs belonging to the W75N (B) massive star-forming region.
    
    \item With our VLA observations we have identified a cluster of at least five YSOs (VLA~1, VLA~2, VLA~3, VLA[SW] and VLA[NE]) in a region of $\sim$10~arsec ($\sim$13000~au). All of the sources for which information was obtained on their structure and SED, exhibit accretion/outflow activity at different relative stages of their evolution. In this sense, given that VLA~1 has indications that it has already started the photoionization stage, it could be relatively more evolved than VLA~2 and VLA~3. To further characterize this cluster of YSOs, ALMA observations with very high-angular resolution are needed to resolve individually the gas/dust content of each YSO, as well as to study their expected different chemical compositions. 
\end{itemize}

We thank our anonymous referee for her/his very valuable comments and suggestions on the manuscript. The work of A.R.-K. and C.C.-G. was supported by Universidad Nacional Aut\'onoma de Mexico DGAPA-PAPIIT grant number IN108218. J.C. acknowledges support from grant PAPIIT-UNAM-IG100218. A.R.-K. thanks the Consejo Nacional de Investigaciones
Cient\'ificas y T\'ecnicas (CONICET) to support postdoctoral research. A.R.-K. and J.M.T. acknowledge support from the European Union's Horizon 2020 Research and Innovation program under the Marie Sklodowska-Curie grant agreement number 734374 - Project acronym: LACEGAL. N.A-L., J.F.G, and J.M.T. are supported by the Spanish grant 
AYA2017-84390-C2-R (AEI/FEDER, UE). S.C. acknowledges support from DGAPA, UNAM and CONACyT, M\'exico.
J.F.G. acknowledges financial support from the State Agency for Research of the Spanish MCIU through the ``Center of Excellence Severo Ochoa'' award for the Instituto de Astrof\'isica de Andaluc\'ia (SEV-2017-0709).

\vspace{1cm}
\noindent \textit{Data availability}\\

\noindent The datasets underlying this article were derived from sources in the public domain: NRAO Data Archive, https://science.nrao.edu/observing/data-archive.











 \appendix
\section{Supplementary tables and figures}\label{appendix}

Table \ref{tbl-images}  lists the parameters of the images used in this paper. The columns are as follows: [1] spectral band, [2] epoch of observation, [3] central frequency, [4] bandwidth, [5] weighting of visibilities (NA = natural, UN = uniform, and R0 and R-1 = Briggs, using robust parameter equal to 0 and -1), [6] synthesized beam size and position angle, [7] rms noise, [8] Figure/Table in which each image is used in the paper.

\begin{figure*}
    \centering
    \includegraphics[width=\textwidth]{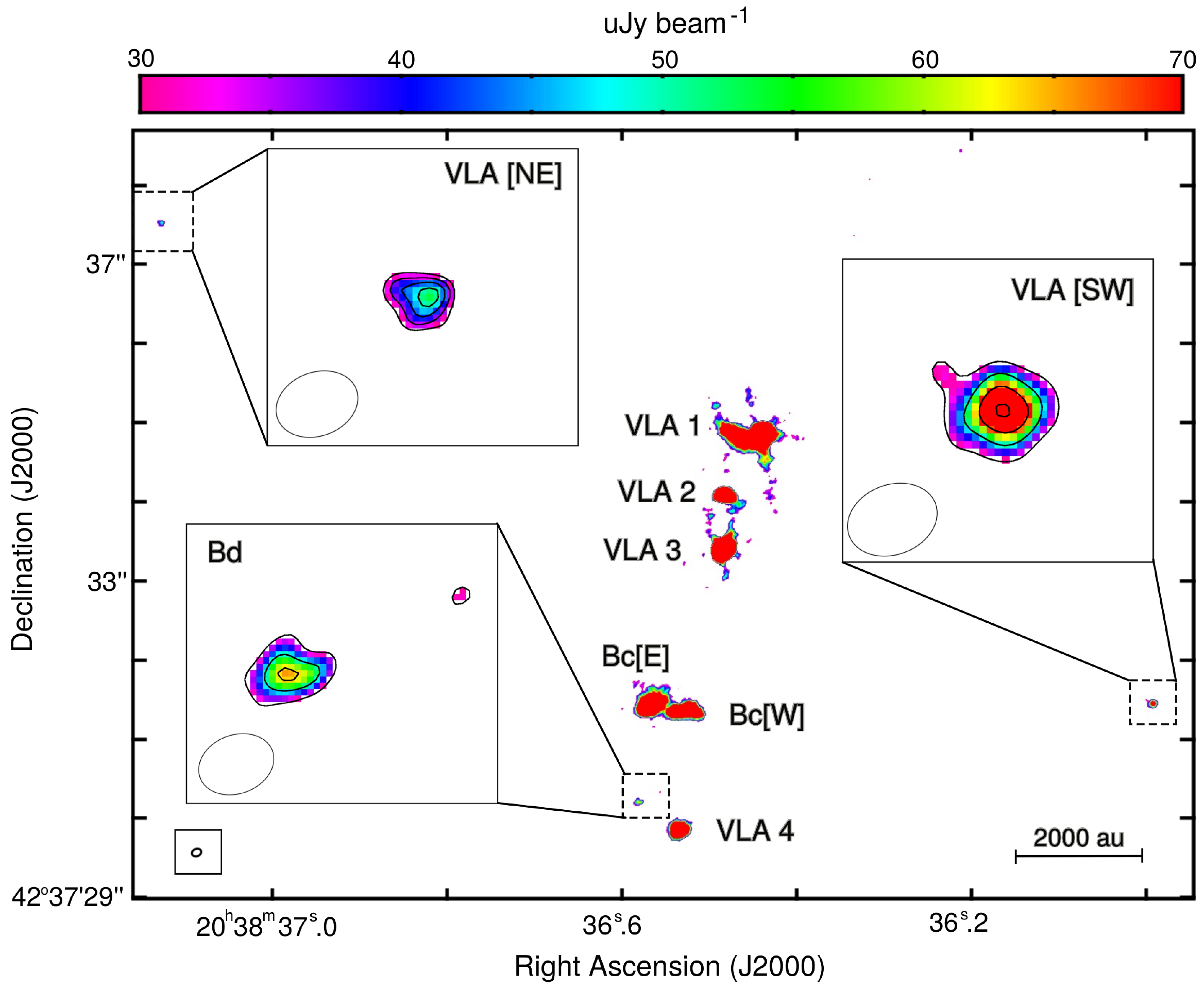}
    \caption{Radio continuum image of W75N~(B) made by the combination of C, Ku, K, and Q bands (epoch 2014.29), using multifrequency synthesis cleaning and Briggs weighting (robust 0). Three new weak ($<$100~$\mu$Jy) compact radio continuum sources are detected: VLA[NE] and VLA[SW] located at $\sim$8 arcsec northeast and $\sim$6 arcsec southwest from VLA 2, respectively, and Bd at $\sim$0.5 arscec northeast from VLA~4. Countour levels correspond to 4, 5, 6, and 7 times 7$\times10^{-6}$ Jy/beam (VLA[NE]);  4, 7, 10, and 13 times 7$\times10^{-6}$ Jy/beam (VLA[SW]); and 4, 7, and 9 times 7$\times10^{-6}$ Jy/beam (Bd). Black rectangles show zoomed-in regions enclosing each of the new sources, where the physical scale is given by the syntesized beam, corresponding to 160~au $\times$ 120~au (PA = --69$^{\circ}$), approximately.
    }\label{new_sources}
\end{figure*}

Fig. \ref{new_sources} shows a radio continuum map of the region, with close-ups of the three new compact sources of $<$ 100 $\mu$Jy detected in the field: VLA[NE], VLA[SW], and Bd.

\begin{table*}
 \center
  \caption{PARAMETERS OF THE VLA IMAGES}
  \label{tbl-images}
    \begin{threeparttable}
  \begin{tabular}{lccccccl}
    \hline
   Spectral Band & Epoch &  Central Frequency  & Total Bandwidth  &  Weighting  & Synthesized &  rms  &  Used in \\
   {}           &  {} &  (GHz)             & (GHz)            &   {}        & Beam        &  ($\mu$Jy/beam) &         {}    \\
    \hline
   X         & 1992.90    &  8.44  &   0.1    &  R0    &  0$\farcs$23 $\times$ 0$\farcs$19; -56$^{\circ}$   & 90  &   Fig.\ref{pm} \\
   Ku        & 2001.31    &  15.0  &   0.1    &  R0    &  0$\farcs$47 $\times$ 0$\farcs$39; -83$^{\circ}$   & 80  &   Fig.\ref{pm}  \\
   X         & 2006.38    &  8.46  &   0.1    &  R0    &  0$\farcs$22 $\times$ 0$\farcs$17; 18$^{\circ}$    & 50  &   Fig.\ref{pm}  \\
   C         & 2014.29    &  6.0   &   4.0    &  UN    &  0$\farcs$28 $\times$ 0$\farcs$20; 83$^{\circ}$    & 30  &   Fig.\ref{bands}, Table \ref{tbl-srcFlux}  \\ 
   Ku        & 2014.29    &  15.0  &   6.0    &  NA    &  0$\farcs$21 $\times$ 0$\farcs$16; -75$^{\circ}$   & 10  &   Fig.\ref{bands}, \ref{VLA1_bandas}, Table \ref{tbl-srcFlux} \\
   K         & 2014.29    &  22.0  &   8.5    &  NA    &  0$\farcs$16 $\times$ 0$\farcs$10; -74$^{\circ}$   & 10  &   Fig.\ref{bands}, \ref{VLA1_bandas}, \ref{pm}, Table \ref{tbl-srcFlux} \\ 
   K         & 2014.29    &  22.0  &   8.5    &  R0    &  0$\farcs$15 $\times$ 0$\farcs$07; -77$^{\circ}$   & 10  &  Table \ref{tbl-vla1ndot} \\ 
   Q         & 2014.29    &  44.0  &   10.0   &  NA    &  0$\farcs$07 $\times$ 0$\farcs$05; -29$^{\circ}$   & 20  &   Fig.\ref{bands}, \ref{VLA1_bandas}, Table \ref{tbl-srcFlux}      \\ 
   Q         & 2014.29    &  44.0  &   10.0   &  UN    &  0$\farcs$04 $\times$ 0$\farcs$05; -9$^{\circ}$    & 100 &   Fig.\ref{VLA1_bandas}\\  
   C+Ku+K+Q  & 2014.29    &  25.9  &   28.5   &  R0    &  0$\farcs$12 $\times$ 0$\farcs$09; -69$^{\circ}$   & 8   &   Fig.\ref{w75n}, \ref{new_sources}, Table \ref{tbl-srcFlux} \\ 
   C         & 2014.29    &  5.0   &   2.0    &  UN    &  0$\farcs$37 $\times$ 0$\farcs$37; 0$^{\circ}$     & 90  &   Fig.\ref{VLA1_sed}     \\
   C         & 2014.29    &  7.0   &   2.0    &  UN    &  0$\farcs$37 $\times$ 0$\farcs$37; 0$^{\circ}$     & 90  &   Fig.\ref{VLA1_sed}\\
   C         & 2014.29    &  7.0   &   2.0    &  UN    &  0$\farcs$27 $\times$ 0$\farcs$20; 87$^{\circ}$    & 50  &   Fig.\ref{fig-vla3}\\
   Ku        & 2014.29    &  13.0  &   2.0    &  NA    &  0$\farcs$37 $\times$ 0$\farcs$37; 0$^{\circ}$     & 90  &   Fig.\ref{VLA1_sed}     \\
   Ku        & 2014.29    &  13.0  &   2.0    &  NA    &  0$\farcs$26 $\times$ 0$\farcs$19; -85$^{\circ}$   & 45  &   Fig.\ref{fig-vla3}   \\
   Ku        & 2014.29    &  13.0  &   2.0    &  UN    &  0$\farcs$17 $\times$0$\farcs$11; 90$^{\circ}$     & 120 &   Fig.\ref{fig-vla3}\\
   Ku        & 2014.29    &  15.0  &   2.0    &  NA    &  0$\farcs$37 $\times$ 0$\farcs$37; 0$^{\circ}$     & 100 &   Fig.\ref{VLA1_sed} \\
   Ku        & 2014.29    &  15.0  &   2.0    &  NA    &  0$\farcs$20 $\times$ 0$\farcs$15; -77$^{\circ}$   & 35  &   Fig.\ref{fig-vla3}\\
   Ku        & 2014.29    &  15.0  &   2.0    &  UN    &  0$\farcs$15 $\times$0$\farcs$09; 89$^{\circ}$     & 100 &   Fig.\ref{fig-vla3}\\
   Ku	     & 2014.29    &  17.0  &   2.0    &  NA    &  0$\farcs$37 $\times$ 0$\farcs$37; 0$^{\circ}$     & 100 &   Fig.\ref{VLA1_sed} \\
   Ku	     & 2014.29    &  17.0  &   2.0    &  NA    &  0$\farcs$18 $\times$ 0$\farcs$14; -66$^{\circ}$   & 30  &   Fig.\ref{fig-vla3} \\
   Ku	     & 2014.29    &  17.0  &   2.0    &  UN    &  0$\farcs$13 $\times$0$\farcs$09; -87$^{\circ}$    & 90  &   Fig.\ref{fig-vla3}\\
   K	     & 2014.29    &  19.4  &   2.0    &  NA    &  0$\farcs$37 $\times$ 0$\farcs$37; 0$^{\circ}$     & 90  &   Fig.\ref{VLA1_sed}   \\
   K	     & 2014.29    &  19.4  &   2.0    &  NA    &  0$\farcs$18 $\times$0$\farcs$11; -74$^{\circ}$    & 20  &   Fig.\ref{fig-vla3}  \\
   K	     & 2014.29    &  19.4  &   2.0    &  UN    &  0$\farcs$13 $\times$0$\farcs$07; -84$^{\circ}$    & 75  &   Fig.\ref{fig-vla3}\\
   K	     & 2014.29    &  21.2  &   2.0    &  NA    &  0$\farcs$37 $\times$ 0$\farcs$37; 0$^{\circ}$     & 100 &   Fig.\ref{VLA1_sed}\\
   K	     & 2014.29    &  21.2  &   2.0    &  NA    &  0$\farcs$17 $\times$0$\farcs$10; -75$^{\circ}$    & 20  &   Fig. \ref{fig-vla3} \\
   K	     & 2014.29    &  21.2  &   2.0    &  UN    &  0$\farcs$12 $\times$0$\farcs$063; -84$^{\circ}$   & 160 &   Fig.\ref{fig-vla3} \\
   K	     & 2014.29    &  22.0  &   2.0    &  NA    &  0$\farcs$37 $\times$ 0$\farcs$37; 0$^{\circ}$     & 150 &   Fig.\ref{VLA1_sed}  \\
   K	     & 2014.29    &  22.0  &   2.0    &  NA    &  0$\farcs$15 $\times$ 0$\farcs$09; -73$^{\circ}$   & 30  &   Fig.\ref{fig-vla3} \\
   K	     & 2014.29    &  22.0  &   2.0    &  UN    &  0$\farcs$10 $\times$0$\farcs$054; -85$^{\circ}$   & 130 &   Fig.\ref{fig-vla3} \\
   K	     & 2014.29    &  23.4  &   2.0    &  NA    &  0$\farcs$37 $\times$ 0$\farcs$37; 0$^{\circ}$     & 120 &   Fig.\ref{VLA1_sed} \\
   K	     & 2014.29    &  23.4  &   2.0    &  NA    &  0$\farcs$15 $\times$ 0$\farcs$10; -74$^{\circ}$   & 20  &   Fig.\ref{fig-vla3}\\
   K	     & 2014.29    &  23.4  &   2.0    &  UN    &  0$\farcs$11 $\times$0$\farcs$056; -84$^{\circ}$   & 90  &   Fig.\ref{fig-vla3} \\
   Q		 & 2014.29    &  41.0  &   2.0    &  NA    &  0$\farcs$37 $\times$ 0$\farcs$37; 0$^{\circ}$     & 220 &   Fig.\ref{VLA1_sed} \\
   Q		 & 2014.29    &  41.0  &   2.0    &  NA    &  0$\farcs$08 $\times$0$\farcs$056; -40$^{\circ}$   & 30  &   Fig.\ref{fig-vla3}  \\
   Q		 & 2014.29    &  41.0  &   2.0    &  R-1   &  0$\farcs$04 $\times$0$\farcs$039; 82$^{\circ}$    & 70  &   Fig.\ref{fig-vla3} \\
   Q		 & 2014.29    &  43.0  &   2.0    &  NA    &  0$\farcs$37 $\times$ 0$\farcs$37; 0$^{\circ}$     & 200 &   Fig.\ref{VLA1_sed} \\
   Q		 & 2014.29    &  43.0  &   2.0    &  NA    &  0$\farcs$07 $\times$ 0$\farcs$05; -27$^{\circ}$   & 30  &   Fig.\ref{fig-vla3}  \\
   Q		 & 2014.29    &  43.0  &   2.0    &  R-1   &  0$\farcs$04 $\times$0$\farcs$039; 7$^{\circ}$     & 80  &   Fig.\ref{fig-vla3}   \\
   Q	     & 2014.29    &  45.0  &   2.0    &  NA    &  0$\farcs$37 $\times$ 0$\farcs$37; 0$^{\circ}$     & 300 &   Fig.\ref{VLA1_sed} \\
   Q	     & 2014.29    &  45.0  &   2.0    &  NA    &  0$\farcs$07 $\times$ 0$\farcs$05; -25$^{\circ}$   & 40  &   Fig.\ref{fig-vla3} \\
   Q	     & 2014.29    &  45.0  &   2.0    &  R-1   &  0$\farcs$04 $\times$0$\farcs$038; -13$^{\circ}$   & 100 &   Fig.\ref{fig-vla3} \\
   Q	     & 2014.29    &  47.0  &   2.0    &  NA    &  0$\farcs$37 $\times$ 0$\farcs$37; 0$^{\circ}$     & 350 &   Fig.\ref{VLA1_sed} \\
   Q	     & 2014.29    &  47.0  &   2.0    &  NA    &  0$\farcs$09 $\times$ 0$\farcs$05; -18$^{\circ}$   & 70  &   Fig.\ref{fig-vla3} \\
   Q	     & 2014.29    &  47.0  &   2.0    &  R-1   &  0$\farcs$06 $\times$0$\farcs$037; -19$^{\circ}$   & 200 &   Fig.\ref{fig-vla3} \\
   Q         & 2014.29    &  40.5  &   1.0    &  R0    &  0$\farcs$05 $\times$0$\farcs$043; -53$^{\circ}$   & 50  &   Fig.\ref{VLA1_sed} \\
   Q         & 2014.29    &  41.5  &   1.0    &  R0    &  0$\farcs$05 $\times$0$\farcs$043; -36$^{\circ}$   & 50  &   Fig.\ref{VLA1_sed}\\
   Q         & 2014.29    &  42.5  &   1.0    &  R0    &  0$\farcs$05 $\times$0$\farcs$042; -24$^{\circ}$   & 55  &   Fig.\ref{VLA1_sed} \\
   Q         & 2014.29    &  43.5  &   1.0    &  R0    &  0$\farcs$05 $\times$0$\farcs$042; -7$^{\circ}$    & 65  &   Fig.\ref{VLA1_sed} \\
   Q         & 2014.29    &  44.5  &   1.0    &  R0    &  0$\farcs$05 $\times$0$\farcs$040; -25$^{\circ}$   & 65  &   Fig.\ref{VLA1_sed}  \\
   Q         & 2014.29    &  45.5  &   1.0    &  R0    &  0$\farcs$05 $\times$0$\farcs$039; -20$^{\circ}$   & 80  &   Fig.\ref{VLA1_sed} \\
   Q         & 2014.29    &  46.5  &   1.0    &  R0    &  0$\farcs$06 $\times$0$\farcs$039; -20$^{\circ}$   & 100 &   Fig.\ref{VLA1_sed} \\
   Q         & 2014.29    &  47.5  &   1.0    &  R0    &  0$\farcs$08 $\times$0$\farcs$040; -16$^{\circ}$   & 180 &   Fig.\ref{VLA1_sed} \\
\hline
  \end{tabular}
  \begin{tablenotes}
      \scriptsize
      \item {}
    \end{tablenotes}
    \end{threeparttable}
 \end{table*}

\bsp	
\label{lastpage}
\end{document}